\documentclass[12pt,preprint,dvips]{aastex}

\usepackage{amsmath}
\usepackage{amssymb}
\usepackage{verbatim}
\usepackage{wasysym}
\usepackage{soul} %% text highlighting, \hl command...
\usepackage{gensymb} %% \degree symbol
\usepackage[dvips]{color}

\newcommand\addtag{\refstepcounter{equation}\tag{\theequation}}

\definecolor{lightgray}{rgb}{.9,.9,.9}

\newenvironment{abstract_itemize}{
\begin{itemize}
  \setlength{\itemsep}{1pt}
  \setlength{\parskip}{0pt}
  \setlength{\parsep}{0pt}
  \setlength{\parindent}{0pt}
}{\end{itemize}}

\usepackage{times}
\usepackage{textcomp}

\usepackage{aas_macros}

\usepackage{booktabs}

\begin{document}

\title{The near-Earth asteroid population from two decades of observations}

\author{Pasquale Tricarico}
\affil{Planetary Science Institute, 1700 East Fort Lowell, Suite 106, Tucson, AZ 85719, USA}
\email{tricaric@psi.edu}

\begin{abstract}
Determining the size and orbital distribution of the population of near-Earth asteroids (NEAs) is the focus of intense research, 
with the most recent models converging to a population of approximately $1000$ NEAs larger than 1~km
and up to approximately $10^9$ NEAs with absolute magnitude $H<30$.
We present an analysis of the combined observations of nine of the leading asteroid surveys over the past two decades,
and show that for an absolute magnitude $H<17.75$,
which is often taken as proxy for an average diameter larger than 1~km, 
the population of NEAs is $920\pm10$, lower than other recent estimates.
The population of small NEAs is estimated at $(4 \pm 1)\times 10^8$ for $H<30$,
and the number of decameter NEAs is lower than other recent estimates.
This population tracks accurately the orbital distribution of recently discovered large NEAs,
and produces an estimated Earth impact rate for small NEAs in good agreement with the bolide data.
\end{abstract}

\maketitle

\begin{center}
\fcolorbox{white}{white}{
\begin{minipage}{\textwidth}
\section*{Highlights}
\begin{abstract_itemize}
\item We estimate the trailing loss effects for nine asteroid surveys
\item We introduce new methods to compute the NEAs search completeness and NEA population
\item We show the effect of commensurabilities between the orbit period of a NEA and the Earth
\item We expect fewer large NEAs as well as fewer decameter NEAs than other models
\item NEA population averaged Earth impact probability is $\sim 4 \times 10^{-9}$~year$^{-1}$ and impact velocity is $\sim 19$~km~s$^{-1}$
\end{abstract_itemize}
\end{minipage}
}
\end{center}

\section{Introduction}

With an observed population exceeding 14,000 bodies, up from a few hundred only two decades ago,
NEAs represent a threat of impact with Earth, as well as targets for robotic and human exploration,
and a potential for in-space resource utilization.
In order to estimate the size frequency and orbital distributions of the NEA population, the techniques typically adopted include
the characterization of the detection efficiency of a reference survey
and subsequent simulated detection of a synthetic population \citep{2004Icar..170..295S,2011ApJ...743..156M},
or the statistical tracking of NEAs from their source regions in the main belt to the inner solar system
and subsequent comparison to the detections by a reference survey \citep{2002Icar..156..399B,2002Icar..158..329M},
or the combination of these two approaches \citep{2016Natur.530..303G}.
The re-detection of NEAs has also been used to estimate the level of completeness in the search of NEAs \citep{2015Icar..257..302H}.
The first approach is typically adopted by single-survey studies which have access to
all the observing data and meta-data and can accurately assess the detection efficiency,
and this has limited in the past the adoption of this approach on a multi-survey scale.
The other approaches require the introduction of weights on the contributions of source regions which are fitted as free parameters,
or bootstrapping procedures in order to generate NEA population close to the observed one.

An opportunity to produce an improved NEA population model
comes from the development of a survey characterization technique which relies only on publicly available observational data.
This technique was recently used to analyze the large volume of observational data (10,033 nights over two decades) from
nine of the most active asteroid surveys \citep{2016AJ....151...80T}, 
to determine their nightly detection efficiency as a function of asteroid apparent magnitude and apparent velocity.
We present the NEA population estimate methods in \S\ref{sec:methods},
including an assessment of the trailing loss,
the role of commensurability with the Earth's orbit period,
and the computation of the Earth impact rate.
Then in \S\ref{sec:results} we present the main results:
the NEA population as a function of absolute magnitude,
the orbital distribution,
and the comparison to bolide data.
Conclusions are briefly drawn in \S\ref{sec:conclusions}.

\section{Methods \label{sec:methods}}

\begin{figure*}
\centering
\includegraphics*[angle=270,width=\columnwidth]{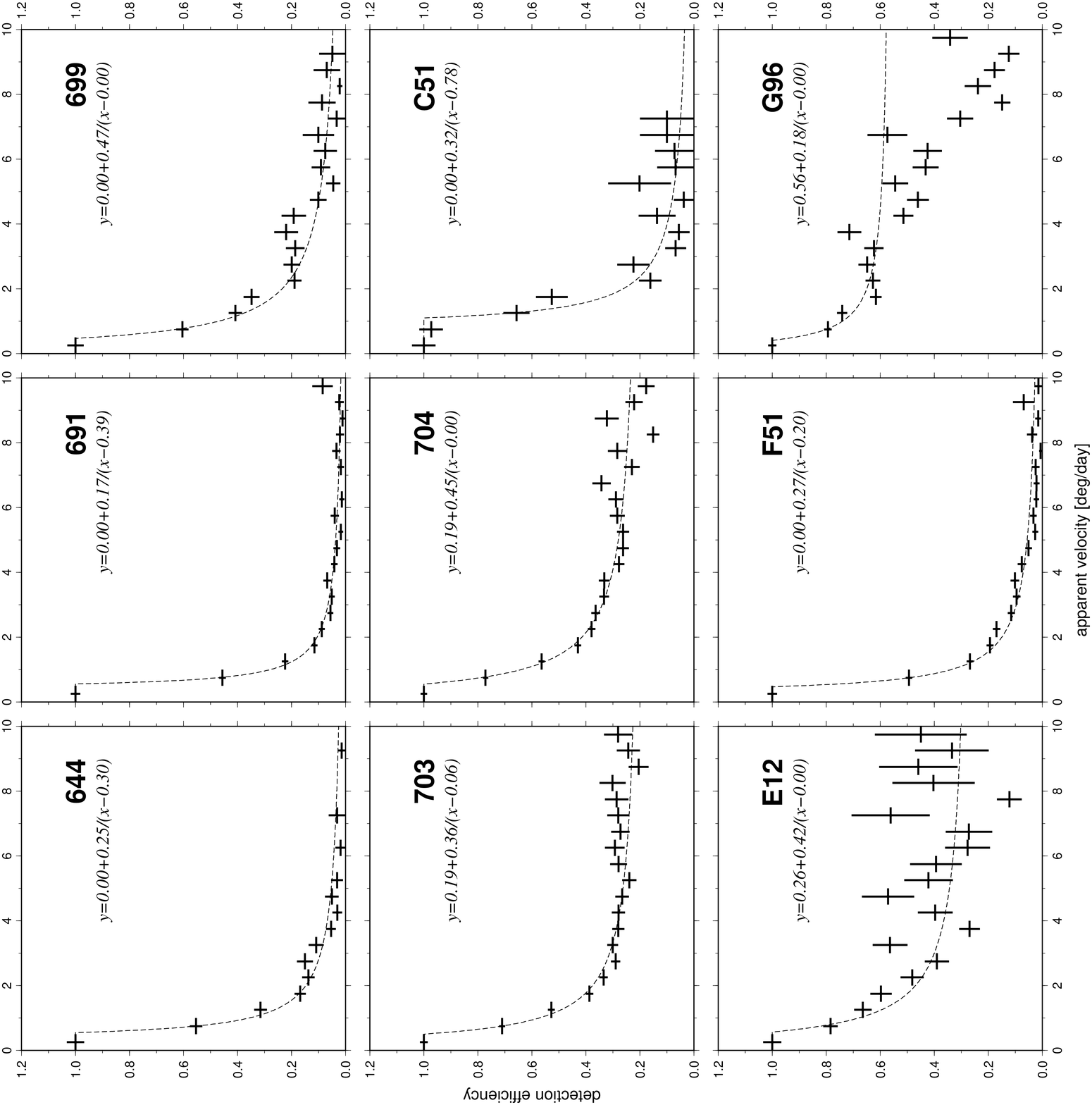}
\caption{
Trailing loss effect as determined from the observations of the nine surveys included,
see \cite{2016AJ....151...80T} for the MPC codes and the properties of each survey.
The data points are the ratio between the number of NEAs observed at a given apparent velocity,
and the expected number of NEAs observed for the same velocity range given the modeled population.
Binning is in increments of 0.5~deg/day,
the first bin is normalized to 1.0,
and the uncertainty is from counting statistics.
}
\label{fig:plot_trailing_grid}
\end{figure*}

The baseline analysis of the observational data \citep{2016AJ....151...80T}
included only observations at apparent velocities up to 100~arcsec/hour,
while NEAs can move at several deg/day, so we need to extend this analysis here
and include the effect of trailing losses.
Trailing losses can be important for asteroids with large apparent velocities,
and this becomes even more important when including very small asteroids which
are only visible when close to the Earth.
In practice, approximately 90\% of all NEAs observations are at apparent velocities below 3~deg/day,
and approximately 99\% below 10~deg/day.
The trailing loss effects can be estimated directly from data,
by comparing the number of asteroids observed at a given range in apparent velocity
with the number expected from the modeled population for the same apparent velocity range,
after accounting for the detection efficiency as modeled in \cite{2016AJ....151...80T}.
The trailing loss effects are estimated separately for each survey,
using a fitting function 
\[ y = c_0 + \frac{c_1}{x-c_2} \addtag \label{eq:fittrail} \]
where the detection efficiency $y=\eta_\text{trail}$ decays as $1/\dot\phi$
with $x=\dot\phi=U$ apparent velocity, see the specific discussion later in this section.
The three free coefficients which are obtained by least-square fitting and allow the function
to be shifted vertically ($c_0$), stretched ($c_1$), and shifted horizontally ($c_2$) in order to better match the observations.
The results for each survey are displayed in Figure~\ref{fig:plot_trailing_grid},
and in general the fit seems satisfying except for G96 for which it was obtained only for points below
3.5~deg/day.
The reason of the difficulty in the G96 fit is unclear, but it may be related to changes in the observing strategies.
Note that since the trailing loss affects the modeled population, and the modeled population affects the trailing loss,
several iterations are necessary before the trailing loss estimate and the modeled population
are stable within a few percents.

\begin{figure*}
\centering
\includegraphics*[angle=270,width=\textwidth]{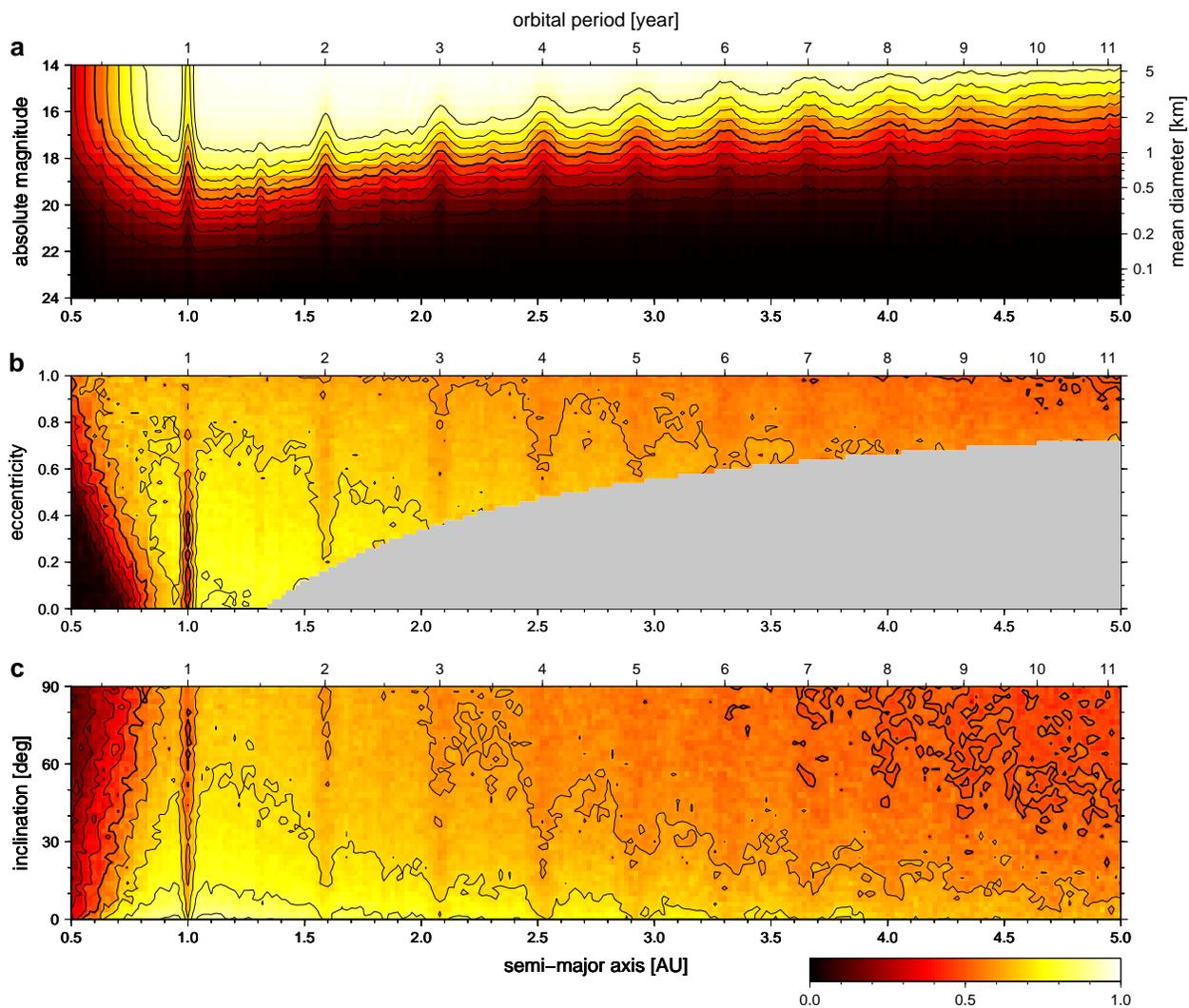}
\caption{
Completeness in the search of NEAs (color scale) projected on:
({\bf a}) absolute magnitude vs.\ semi-major axis, averaged over orbits with eccentricity between 0 and 1,
and inclination between 0\degree{} and 90\degree,
with a mean diameter assuming a nominal geometric albedo $p_V=0.14$;
({\bf b}) eccentricity vs.\ semi-major axis for NEAs with absolute magnitude $14<H<24$,
averaged over inclination between 0\degree{} and 90\degree;
({\bf c}) inclination vs.\ semi-major axis for NEAs with absolute magnitude $14<H<24$,
averaged over orbits with eccentricity between 0 and 1.
Note the effect of commensurability in orbit period of NEA and Earth,
not just the integer multiples but also fractional ones.
}
\label{fig:plot_completeness_multi}
\end{figure*}

The progress in searching for NEAs is tracked by generating $\sim 10^6$ synthetic NEA orbits,
and then numerically propagating the orbits over the two decades period considered, while checking against the sky
coverage and detection efficiency of the surveys \citep{2016AJ....151...80T}.
The absolute magnitude of each synthetic NEA is randomly sampled using $\sim 10^2$ values within $8<H<30$.
For every night $j$ in which the synthetic asteroid appears in the field of view of a survey,
a detection efficiency $\eta_j(V,U)$ can be associated to it, as a function
of apparent magnitude $V$ and apparent velocity $U$ \citep{2016AJ....151...80T}.
We can then calculate the completeness $C$ of the search of the synthetic NEA as 
\[ C = 1 - \prod_j (1-\eta_j) \ . \addtag \]
This relation can be derived considering that
the probability of \emph{not} being detected the night $j$ is $1-\eta_j$,
the product over $j$ indicates the probability of not being detected any of the $j$ nights,
so that $C$ is the probability of being detected at least on one night.
Both $\eta_j$ and $C$ take values between 0 and 1.
The completeness $C$ is larger than or equal to the largest $\eta_j$ and never decreases as nights accumulate.
Note how the completeness defined here is different from the bias $B$ defined in \cite{2016Icar..266..173J} and used in \cite{2016Natur.530..303G},
with $B=\sum_j\eta_j$ where $j$ is each field of view, so $B$ is not bound and can grow larger than 1.0
when enough nights are included, and this requires them to track not only the number of observed objects, but also how many times each object has been observed.
As we show in Figure~\ref{fig:plot_completeness_multi}, some orbits can be significantly 
more difficult to search due to close commensurability with the Earth's orbit period,
which causes the persistency of unfavorable observing geometries.
We note that to our knowledge this is the first published work to clearly show this effect.
In general, high inclination orbits are more difficult to search,
as well as high eccentricity orbits for semi-major axis larger than 1~AU,
while low eccentricity orbits are more difficult to search for semi-major axis up to 1~AU.

\begin{figure*}
\centering
\includegraphics*[angle=270,width=0.7\textwidth]{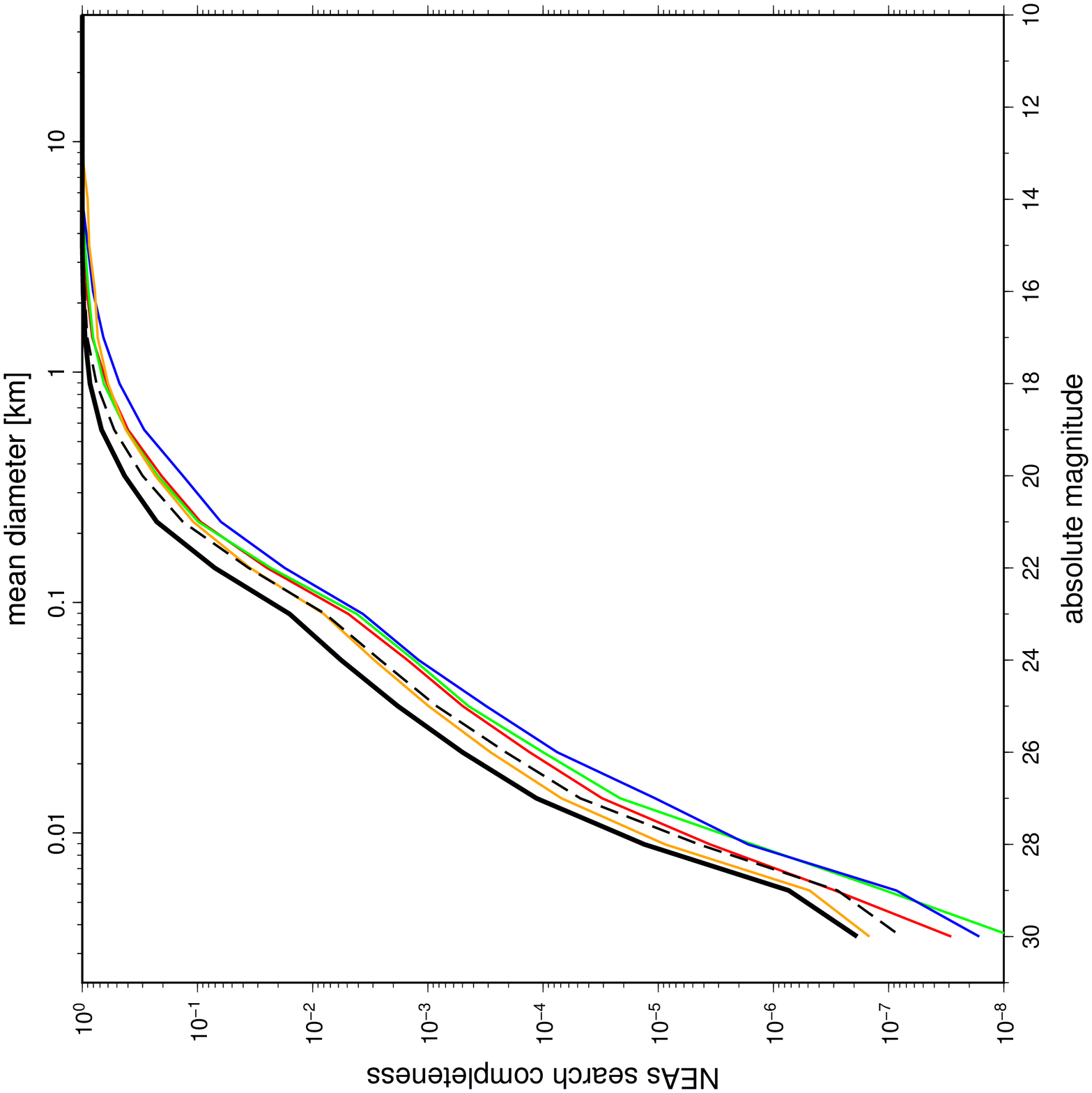}
\caption{
Completeness in the search of NEAs as a function of absolute magnitude $H$.
The bold line is for the full baseline analysis,
while the color lines are for individual surveys
(red=703, green=704, blue=F51, orange=G96).
The dashed line is for data before 2010.
The \emph{dip} at $H=23$ is due to the switching of binning from fine to coarse.
The mean diameter is for reference and assumes a nominal geometric albedo $p_V=0.14$.
}
\label{fig:plot_C_comparison}
\end{figure*}

\cite{2015Icar..257..302H} propose a method to estimate the slope of the completeness curve at large $H$,
assuming that the apparent rate of motion $\dot\phi$ depends on the distance $\Delta$ between an asteroid and the Earth as $\dot\phi\propto 1/\Delta$.
However, if we consider the passage of an asteroid moving on a straight line at constant velocity $v$ relative to the Earth,
the distance $\Delta$ between the asteroid and the Earth follows the relation
$\Delta^2=\Delta_0^2+v^2 (t-t_0)^2$
with $\Delta_0$ the minimum distance reached at time $t_0$,
and in general the apparent rate of motion is 
\[ \dot\phi= \frac{v \Delta_0}{\Delta^2} \addtag \label{eq:alphadot} \]
with $\phi=\arctan[v (t-t_0)/\Delta_0]$.
Continuing along the line of \cite{2015Icar..257..302H} but with this updated formula for $\dot\phi$,
the expected signal-to-noise ratio $S/N$ is
\[S/N \propto \frac{10^{-0.4 H}}{\dot\phi \Delta^2} \propto \frac{10^{-0.4 H}}{v \Delta_0} \addtag \]
and in general the dependence on $\Delta$ is simplified away,
but we can still bracket the slope of the completeness by looking at the limit cases:
\begin{itemize}
\item in the limit of slowly varying $\dot\phi$ at large $\Delta$,
a constant $S/N$ leads to $\Delta^2 \propto 10^{-0.4 H}$ and the completeness $C\propto \Delta^2 \propto 10^{-0.4H}$;
\item when $\Delta\simeq\Delta_0$ we recover the original result of \cite{2015Icar..257..302H} where $\dot\phi\simeq v/\Delta$,
and a constant $S/N$ leads to $\Delta \propto 10^{-0.4 H}$ and the completeness $C\propto \Delta^2 \propto 10^{-0.8H}$.
\end{itemize}
What we see in Figure~\ref{fig:plot_C_comparison} is that
the completeness curves decrease as $C\propto 10^{-0.4H}$ at $H\simeq 20$, 
and then become steeper as $H$ increases, 
with $C\propto 10^{-0.6H}$ at $H\simeq 26$,
and then reaching $C\propto 10^{-0.8H}$ or steeper at $H \geqslant 27$.
The gradual slope increase seems to indicate the transition from a regime 
dominated by the observation of slow $H\simeq 20$ objects away from their minimum Earth approach distance,
to one dominated by fast $H \geqslant 27$ objects observed mostly during their close Earth flyby.
A completeness curve dropping steeper than $C\propto 10^{-0.8H}$ at large $H$
may indicate a limit of applicability of the simplified approach of \cite{2015Icar..257..302H}.
Note however that since G96 is the main contributor to the completeness curve at large $H$,
the difficulty in the corresponding trailing loss fit (see Figure~\ref{fig:plot_trailing_grid})
may play a role in observing the completeness curve dropping steeper than $C\propto 10^{-0.8H}$.

\begin{figure*}
\centering
\includegraphics*[angle=270,width=0.6\textwidth]{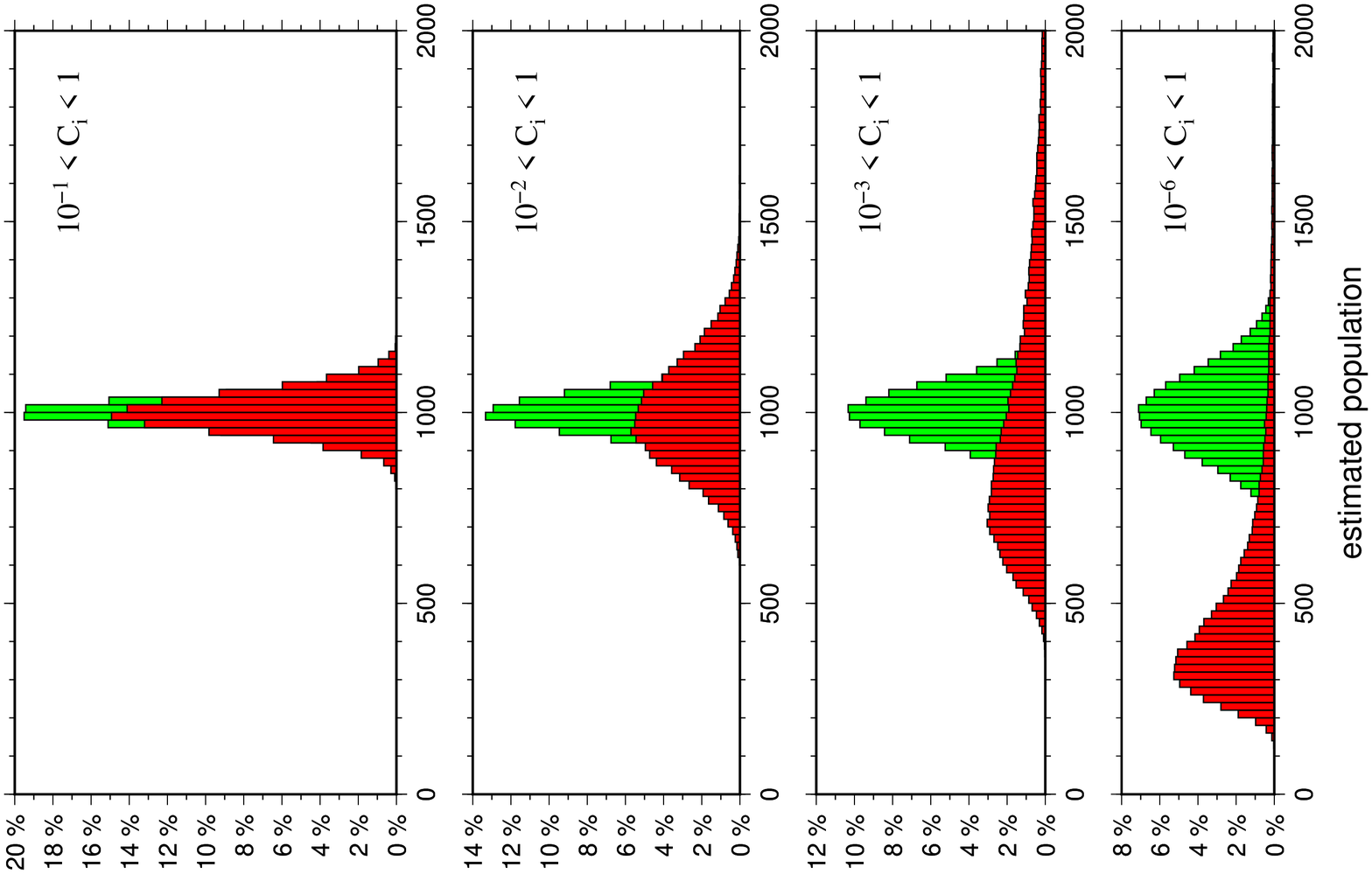}
\caption{
Numerical validation of the negative binomial approach.
Each panel shows the distribution of the estimated population over a large number of simulations,
using completeness values $C_i$ sampled within the range indicated in each panel (see main text).
The population has been estimated by applying the negative binomial statistics bin-by-bin (red),
and by merging bins (green).
As the dynamical range of $C_i$ increases, the bin-by-bin estimate shows a bias towards smaller values,
while the merged bins population estimate remains stable around the nominal population value of $10^3$.
}
\label{fig:plot_binomial_tests}
\end{figure*}

The orbits of the synthetic NEAs are sampled as follows:
semi-major axis from 0.5~AU to 5.0~AU;
eccentricity from 0 to 1;
inclination from 0\degree\ to 90.0\degree;
longitude of the ascending node, argument of perihelion, and mean anomaly from 0\degree\ to 360\degree.
By definition, all NEAs have orbits with a perihelion distance smaller than 1.3~AU.
The absolute magnitude $H$ is sampled from 8 to 30.
The orbits and the corresponding completeness values are then binned,
and the completeness $C_i$ associated to each bin $i$ is simply the average value.
The bin population $n_i$ is estimated using the negative binomial (NB) statistics,
with $k_i$ the number of known NEAs observed in the same bin.
The probability for each possible value of $n_i \geqslant k_i$ is
\[ P_\text{NB}(n_i,k_i,C_i) = \frac{(n_i-1)!}{(k_i-1)!(n_i-k_i)!} C_i^{k_i} (1-C_i)^{n_i-k_i} \addtag \label{eq:negbin} \]
with mean $n_i=k_i/C_i$ and variance $\sigma^2_{n_i}=k_i(1-C_i)/C_i^2$.
The NB statistics is a powerful tool but produces a population estimate $n_i\geqslant1$ only for bins with $k_i\geqslant1$,
meaning that all empty bins will not contribute to the overall population,
independently of their completeness value $C_i$.
To understand how this affects the population estimate,
we have performed several numerical tests, and for each test:
\begin{itemize}
\item we pick a range for $C_i$ so that $10^{-b}<C_i<1$ with $b$ regulating the dynamic range;
\item assign randomly generated $C_i$ values to $10^5$ bins, using a distribution which is uniform in logarithmic scale, in order to have approximately a constant number of entries per decade;
\item assign randomly $10^3$ objects to the $10^5$ bins (the population $n_i$ that we want to estimate);
\item sample $k_i$ using binomial statistics (the observed objects with $0\leqslant k_i\leqslant n_i$);
\item the $10^5$ pairs of $C_i$ and $k_i$ values are fed into NB statistics to estimate $n_i=k_i/C_i$.
\end{itemize}
The results, displayed in Figure~\ref{fig:plot_binomial_tests},
seem to indicate that the problem of empty bins is moderate if the dynamic range of $C_i$ is limited to one or two orders of magnitudes,
but becomes important as the dynamic range of $C_i$ increases, producing total population estimates systematically lower than nominal. 
A solution to this issue is that of estimating the population for a large group of bins at once,
instead of estimating the population bin-by-bin.
If we replace the per-bin variables $n_i$ and $k_i$
with the total values $n_\text{tot}$ and $k_\text{tot}$ over the group of bins considered,
and replace $C_i$ with the average value $\langle C_i \rangle$ over the bins considered,
we can then apply the NB formalism, and as Figure~\ref{fig:plot_binomial_tests} shows,
the estimates are consistent with the nominal value even for very large dynamic range.
One corollary of this is that since there is no general recipe for grouping bins,
the population estimate will depend on how bins are grouped,
or equivalently on how the orbital elements volume is sliced and diced in the case of NEAs.
We find that it is important for the grouping to take into account the global variations of the 
completeness (see Figure~\ref{fig:plot_completeness_multi})
as well as the dependence on the orbital elements.
So we use two different sets of binning:
a fine binning with a bin size of 
0.05~AU in semi-major axis,
0.05 in eccentricity,
$5^\circ$ in inclination,
and 1~magnitude in $H$;
and a coarse binning where we keep the same bin size for semi-major axis and $H$,
while we merge all bins in eccentricity and inclination.
The fine binning is used for $H<23$ and above that we use the coarse binning,
as the two population estimates overlap at $H\simeq23$.

The impact rate of synthetic NEAs with Earth is computed using standard techniques \citep{1981Icar...48...39K}
accurately validated on known test cases \citep{1998A&AS..133..437M},
and includes an effective Earth cross section $\sigma_E=\pi R_E^2 (1+v_\text{esc}^2/v_\infty^2)$ enhanced by gravitational focusing,
with $R_E$ Earth radius, 
$v_\text{esc}$ Earth escape velocity,
and $\vec{v_\infty}=\vec{v_\text{ast}}-\vec{v_\text{E}}$ the orbital intersection velocity
of the asteroid relative to Earth ignoring the Earth's gravitational pull.
The distribution of $v_\infty$ is typically wide and irregular \citep{1994Icar..107..255B},
but averaging over the mean values within a bin is sufficient in the central limit for a sufficiently large number of samples.
The Earth impact velocity is then $v_\text{imp}^2=v_\text{esc}^2+v_\infty^2$.

\section{Results \label{sec:results}}

\begin{figure*}
\centering
\includegraphics*[angle=270,width=0.7\textwidth]{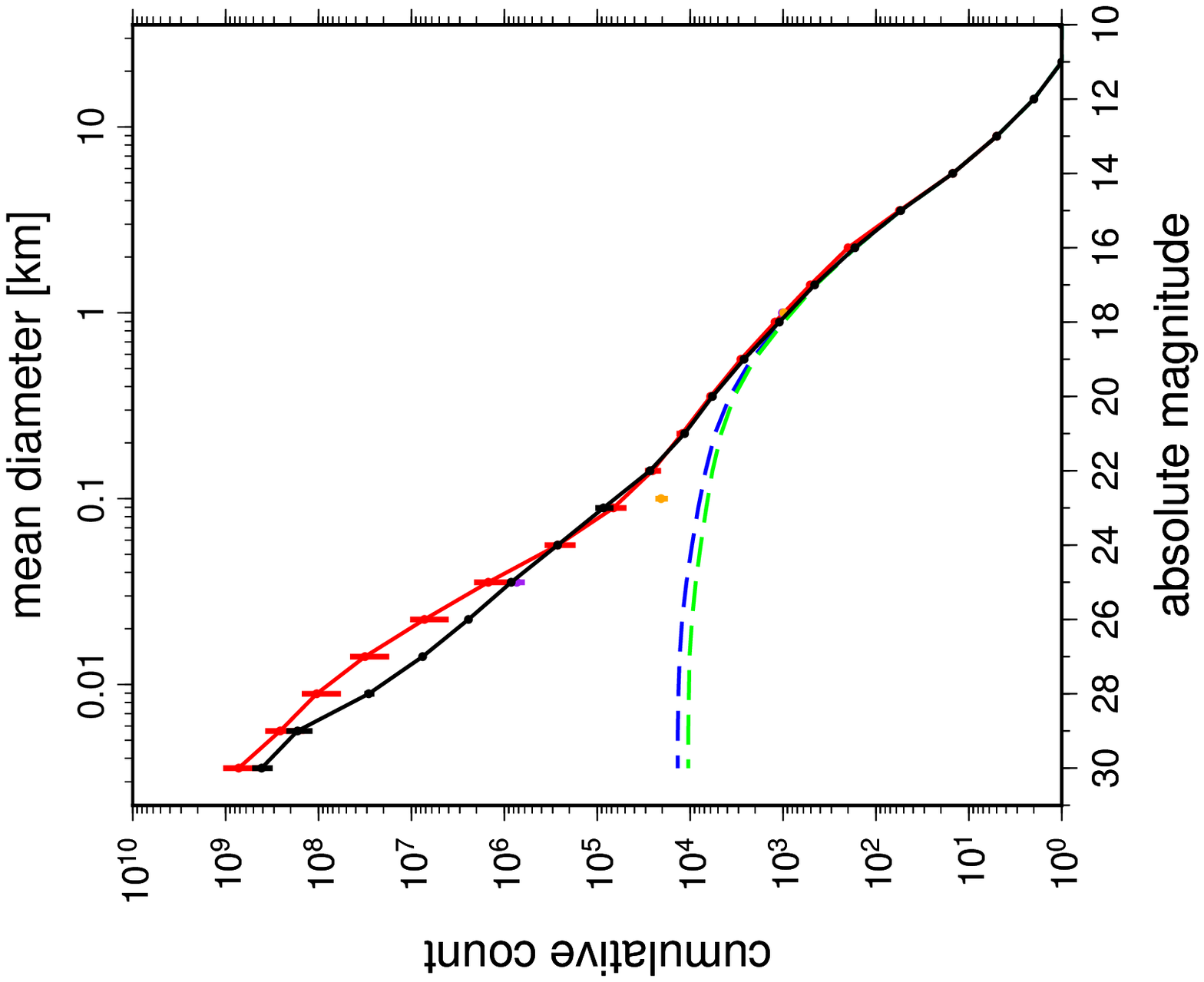}
\caption{
Cumulative NEA population as a function of absolute magnitude.
The black curve is our estimated population,
where the error bars are present but small,
see Table~\ref{tab:pop_data}.
The orange data are from \cite{2011ApJ...743..156M},
the red from \cite{2015Icar..257..302H},
the purple from \cite{2016Natur.530..303G}.
The dashed green data are the observed NEA population
used as input (observed during the baseline period up to August 2014),
while the blue data are the full known NEA population as of February 2016.
The mean diameter is for reference and assumes a nominal geometric albedo $p_V=0.14$.
}
\label{fig:plot_cumulative_pop}
\end{figure*}

The resulting population debiasing scheme produces a cumulative NEA population model as a function of absolute magnitude,
see Table~\ref{tab:pop_data} and Figure~\ref{fig:plot_C_comparison}
where it is displayed along with previous results by other studies.
We estimate $920\pm10$ NEAs with absolute magnitude $H<17.75$,
roughly corresponding to asteroids larger than 1~km at a nominal average geometric albedo $p_V=0.14$ \citep{2004Icar..170..295S}.
This is lower than other recent results \citep{2011ApJ...743..156M,2015Icar..257..302H,2016Natur.530..303G}
that converge to approximately 1000 large NEAs.
We also estimate $(7 \pm 2)\times 10^4$ NEAs with absolute magnitude $H<22.75$,
roughly corresponding to asteroids larger than 100~m,
in good agreement with \cite{2015Icar..257..302H} and \cite{2016Natur.530..303G}
but higher than the estimate from NEOWISE \citep{2011ApJ...743..156M}.
The data available allow to estimate the small NEA population for $H<30$ at $(4 \pm 1)\times 10^8$.
The side-by-side comparison with \cite{2015Icar..257..302H} in Figure~\ref{fig:plot_cumulative_pop} indicates a generally good agreement,
except for a possible decrease in the number of decameter NEAs.

\begin{figure*}
\centering
\includegraphics*[angle=270,width=0.7\textwidth]{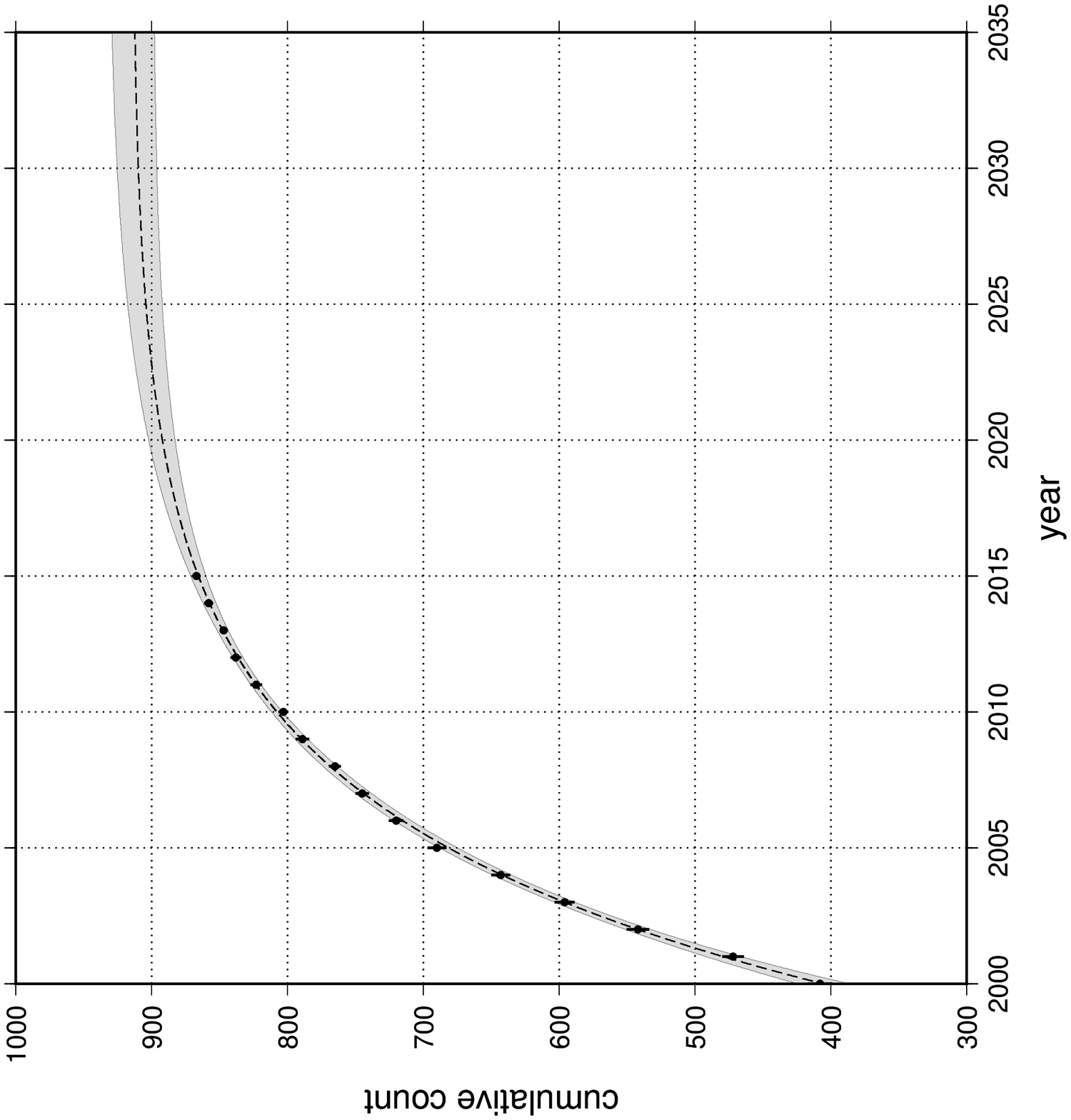}
\caption{
Cumulative number of discovered NEAs with $H<17.75$ as a function of time,
between the years 2000 and 2015 included.
A least squares fit of the data produces an extrapolated total number of $914^{+19}_{-15}$ NEAs.
The nominal uncertainty assigned to each point is the square root of the number of discoveries in that year.
}
\label{fig:plot_discovery}
\end{figure*}

As we show in Figure~\ref{fig:plot_discovery} we can empirically support the result of a lower number of large NEAs
by looking at the cumulative number of discovered NEAs with $H<17.75$ as a function of time.
By using a function that tends exponentially to the total number of $H<17.75$ NEAs,
we can least squares fit the data obtaining an extrapolated total population of $H<17.75$ NEAs of $914^{+19}_{-15}$ (1 sigma),
in good agreement with the result of $920\pm10$ from the full analysis.
While we acknowledge that such a test can only be used empirically,
it seems to confirm that the total number is significantly lower than 1000,
and if one were to introduce corrections to account for the increasing performance 
of surveys over time, this would further strengthen the results of this test.

\begin{figure*}
\centering
\includegraphics*[angle=270,width=0.7\textwidth]{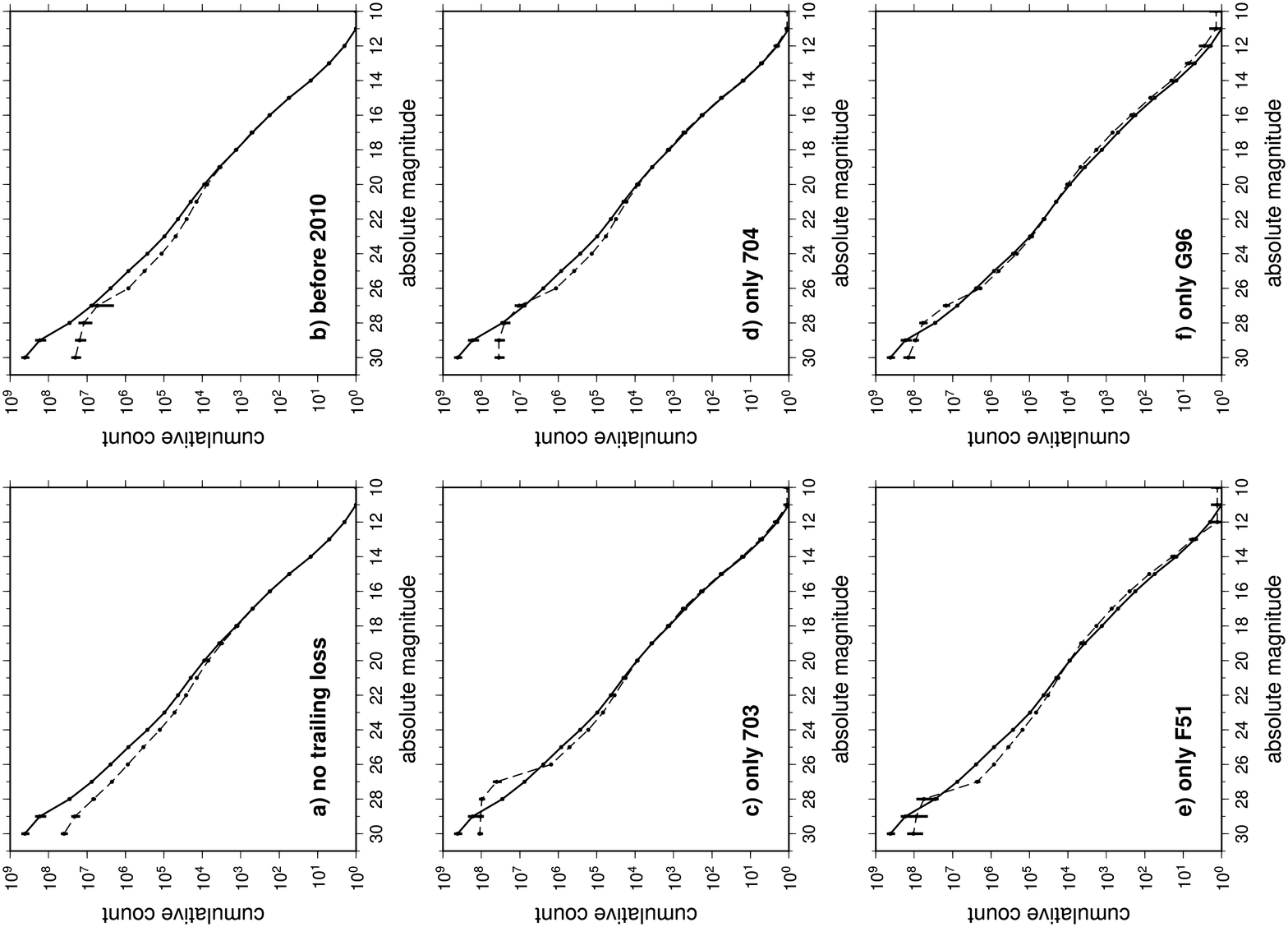}
\caption{
Cumulative NEA population as a function of absolute magnitude.
In each panel the solid curve is our estimated population,
while the dashed curve is the population under different conditions:
({\bf a}) excluding the NEAs trailing loss effects;
({\bf b}) using only observations up to December 2009 included;
({\bf c}) using only observations by (703) Catalina Sky Survey;
({\bf d}) using only observations by (704) LINEAR;
({\bf e}) using only observations by (F51) Pan-STARRS;
({\bf f}) using only observations by (G96) Mt.~Lemmon Survey;
}
\label{fig:plot_pop_comparison}
\end{figure*}

We have performed an analysis of the stability and sensitivity of the produced population with respect to the trailing loss modeling,
to a drastic reduction in the input observations,
and to observations from single surveys, see Figure~\ref{fig:plot_pop_comparison}.
Trailing loss produces a progressively large correction as we move to large $H$,
from a negligibly small correction $H\leqslant20$ up to a full order of magnitude correction at $H\simeq30$,
because the observed trailing loss is very strong, see Figure~\ref{fig:plot_trailing_grid}.
If we include observations up to December 2009 included,
corresponding to including only approximately 60\% of the observed NEAs compared to the full period,
we still obtain a remarkably good agreement with the full population for $H\leqslant20$,
and then some deviations above that.
We have a similar level of agreement when we including only observations from a single survey.
The deviations are most likely of statistical nature,
due to the use of subsets of asteroids and observing nights as model inputs.
An additional contribution to deviations for single survey population estimates may be due to follow-up observations:
when a survey interrupts normal search activities to confirm a NEA just discovered by another survey,
in a sky region relatively far from the previous search area, 
that biases the single survey estimate as if that NEA was discovered by the follow-up survey,
causing a bump in the population estimate.
By looking back at Figure~\ref{fig:plot_cumulative_pop} we can get a feeling of the
difficulty in assessing assess whether the differences
with \cite{2015Icar..257..302H} are statistically significant or not,
since these differences have magnitudes comparable with that of the test cases in Figure~\ref{fig:plot_pop_comparison}.

\begin{figure*}
\centering
\includegraphics*[angle=270,width=\textwidth]{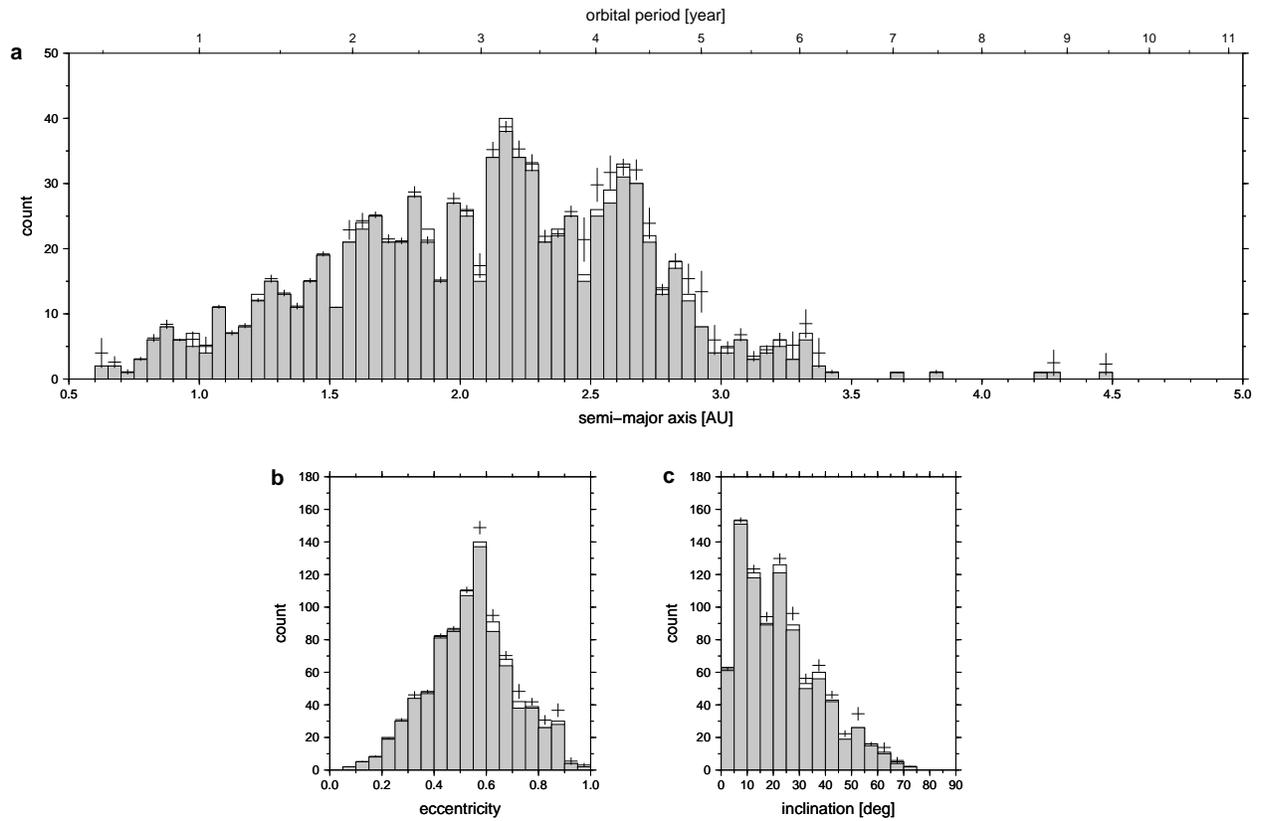}
\caption{
Distribution of NEAs with $H<17.75$ versus ({\bf a}) semi-major axis, ({\bf b}) eccentricity, ({\bf c}) inclination.
The histograms represent the observed population (updated to February 2016),
with diagonal filling for NEAs discovered during the last 18 months,
while the error bars represent the modeled population.
}
\label{fig:plot_pop_histo_1775}
\end{figure*}

Our NEA population model can be tested against the current discoveries of large NEAs.
In Figure~\ref{fig:plot_pop_histo_1775} we show the orbital distribution of the $H<17.75$ population.
As we can see, the estimated population, which used input observations up to August 2014,
predicts closely the current observed population including the NEAs discovered during the following 18 months
up to February 2016.
At higher magnitudes, the orbital distribution does not appear to depend on $H$,
with the only possible exception of a decrease of the effective maximum inclination of NEAs at increasing $H$,
but more data are required for a statistically significant analysis of the orbital distribution at large $H$.

\begin{figure*}
\centering
\includegraphics*[angle=270,width=\textwidth]{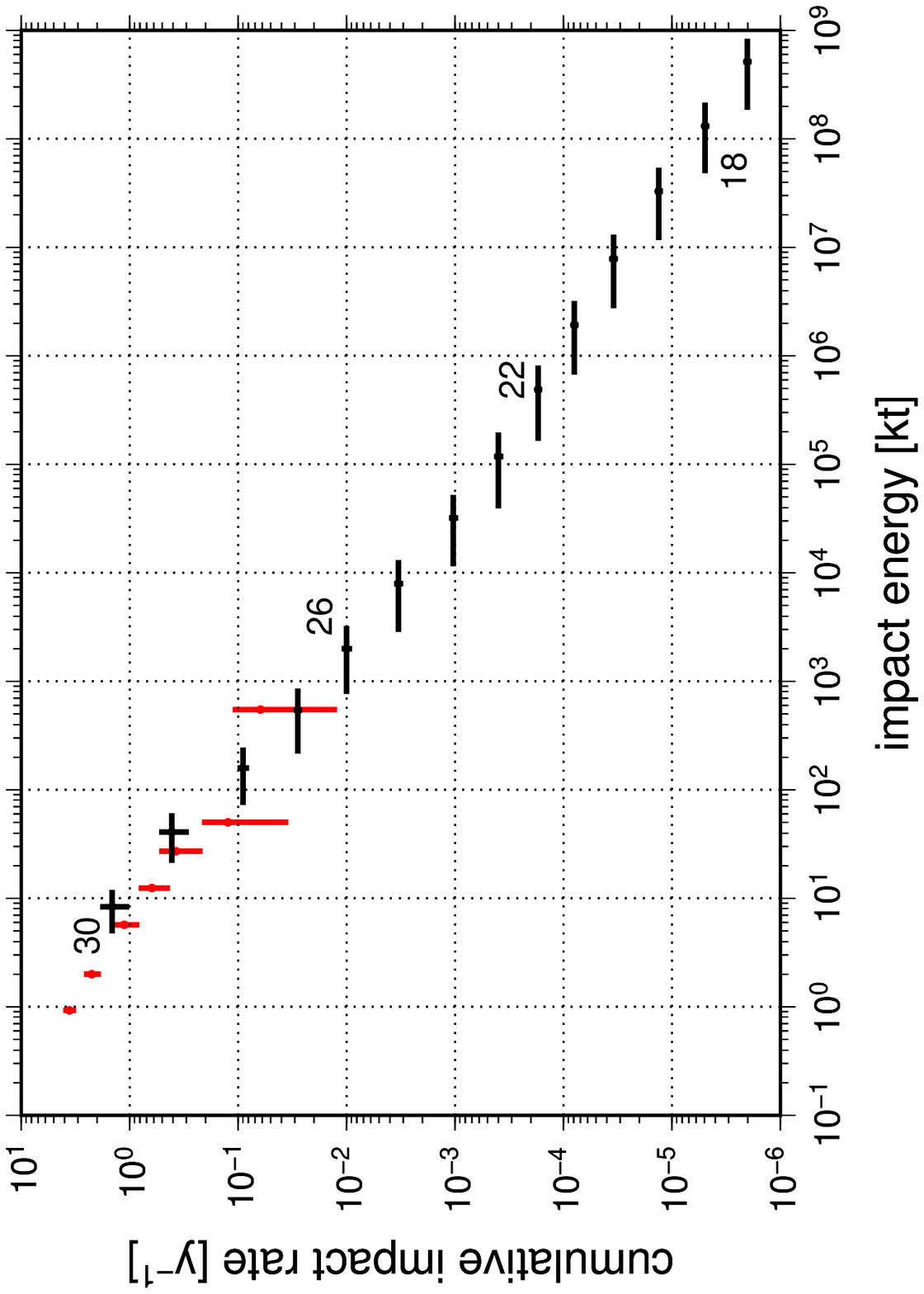}
\caption{
Cumulative Earth impact rate for NEAs as a function of the impact energy (black),
in increments of one absolute magnitude $H$, with the top left point corresponding to $H<30$ as indicated.
The bolide data \citep{2013Natur.503..238B} is in red (error bars are counting statistics only).
The assumed mean density of NEAs is 3,000~kg~m$^{-3}$ and the geometric albedo $p_V=0.14$.
The impact energy is in kiloton (1 kt = $4.184 \times 10^{12}$~J),
and the impact energy uncertainty includes contributions from the impact velocity range as well as the size range in each magnitude bin.
}
\label{fig:plot_cumulative_pop_energy_rate}
\end{figure*}

\begin{figure*}
\centering
\includegraphics*[angle=270,width=\textwidth]{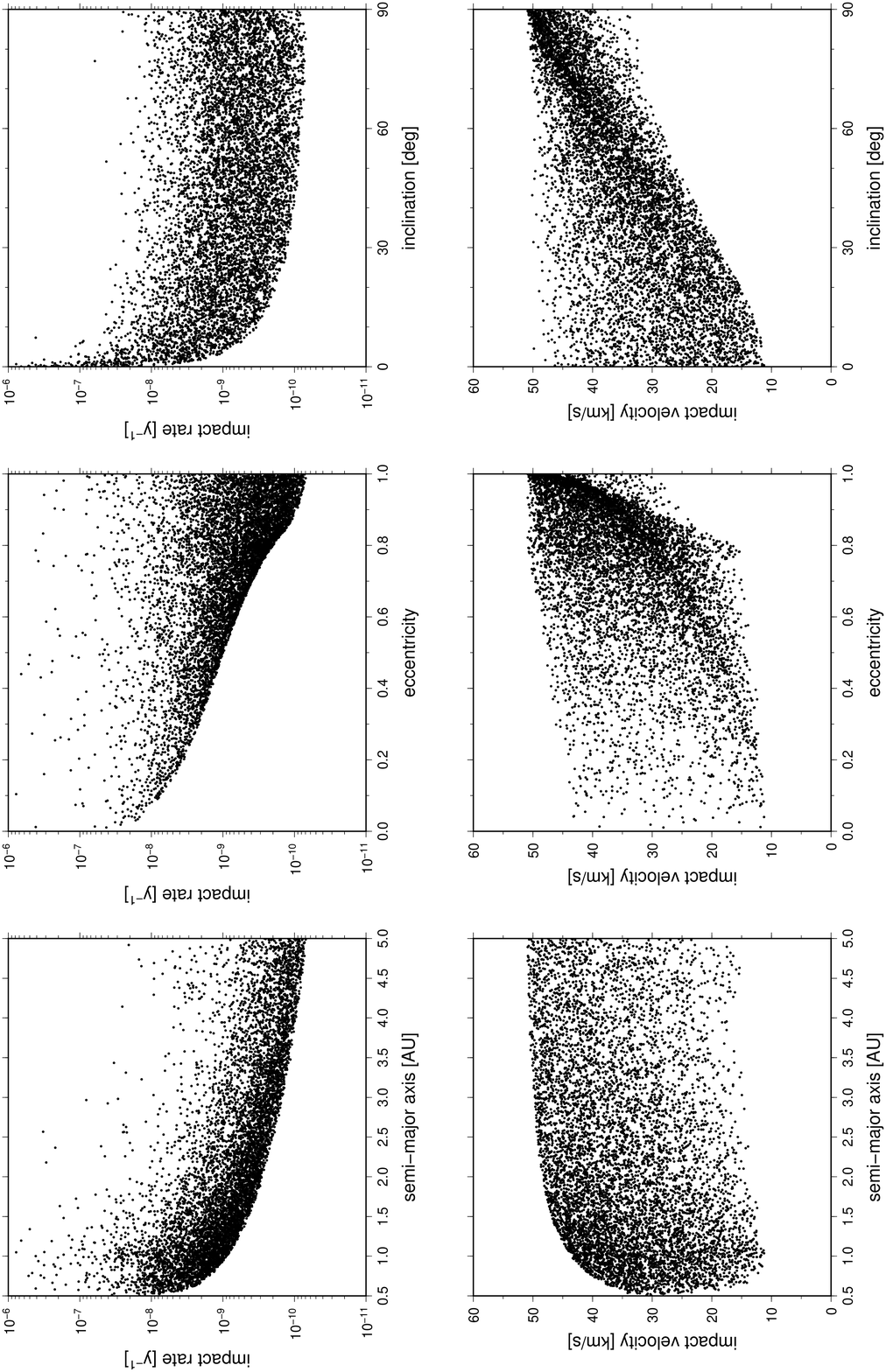}
\caption{
Earth impact rate (top row) and impact velocity (bottom row) for synthetic NEAs as a function of their uniformly sampled orbital elements.
Note: approximately 23\% of the orbits have no impact solutions and are not plotted;
these non-impacting orbits have $q>Q_\text{E}$ or $Q<q_\text{E}$,
with $q$ and $Q$ respectively perihelion and aphelion distance,
and E for Earth.
}
\label{fig:plot_collision}
\end{figure*}

This NEA population can also be tested against the observed rate of bolides \citep{2013Natur.503..238B},
and in Figure~\ref{fig:plot_cumulative_pop_energy_rate} we compare
their impact rate versus impact energy with ours.
Our modeled NEAs impact rates overlap the bolide observations,
and the two datasets also appear to have the same slope.
The uncertainty in bolide data represents only counting statistics,
and there could be some systematic uncertainties in some of the model assumptions\
(i.e.\ from mean optical to total energy conversion)
capable of producing a shift of about a factor two
(P.~Brown, \emph{priv.\ comm.}).
Note that in the calculation of the impact probabilities,
earlier studies \citep{2013Natur.503..238B,2015Icar..257..302H}
used a single average value for the intrinsic impact probability with Earth of $2\times 10^{-9}$~year$^{-1}$,
while we have calculated it explicitly for every NEA population bin.
As we show in Figure~\ref{fig:plot_collision}
the impact probability can take values over several orders of magnitude,
and it would be difficult to assume that a single value is representative of this wide range.
For the purpose of comparison with previous works we can still obtain an average value 
of approximately $6 \times 10^{-9}$~year$^{-1}$ when giving equal weight to all orbits sampled,
or approximately $4 \times 10^{-9}$~year$^{-1}$ when weighting orbits by the estimated population.
These rates are slightly larger than what has been recently used in the literature.
Similarly we can provide some nominal average values for the Earth impact velocity,
which takes values between the Earth escape velocity of $11.2$~km~s$^{-1}$ up to approximately $51$~km~s$^{-1}$,
and for equal weight orbits we obtain an average impact velocity of approximately $35$~km~s$^{-1}$,
while by weighting orbits by their Earth impact rate we obtain an average impact velocity of approximately $19$~km~s$^{-1}$,
with a similar value if we also weight by the estimated population, in line with commonly used values.

\section{Conclusions \label{sec:conclusions}}

We have used the combined observations of asteroid surveys over the past two decades
to estimate the trailing loss effect NEAs at fast apparent velocities,
to determine the current level of completeness in the search of NEAs,
and to model the NEA population.
The resulting NEA population estimate is in general agreement with other recent models,
except for a possible decrease in the number of large NEAs as well as a lower number of decameter NEAs.
The population estimate methods produce consistent results when applied subsets of all the available observations,
track very accurately the orbital distribution of the large NEAs being discovered over the past 18 months,
and lead to bolide rate estimates in line with current observations.
Additionally, the methods employed highlight the role of orbital commensurabilities with the Earth's orbit in the search completeness.

\acknowledgments

This research is supported by the NASA Near Earth Object Observations (NEOO) Program grant NNX13AQ43G,
and by the NASA Applied Information System Research (AISR) Program grant NNX08AD18G.
We gratefully acknowledge the referees Alan Harris and Mikael Granvik
for their detailed reviews which led
to the identification and resolution of several issues 
and to a significant overall improvement of the manuscript.
We thank Peter Brown for correspondence on bolide observations,
and Bill Bottke and Robert Jedicke for discussions of results in earlier versions of this manuscript.
Early versions of this research made use of the BOINC-based \citep{BOINC} orbit@home distributed computing network,
and the participants who donated computing time on their personal computers are acknowledged.

\begin{deluxetable}{crrrcr}
\tablewidth{0pt}
\tablecaption{Cumulative NEA population}
\tablehead{ \colhead{$H$} & \colhead{observed} & \colhead{known} & \multicolumn{3}{c}{estimated population} }
\startdata
10 & $1$ & $1$ & $1.0$ & $\pm$ & $0.0$\\
11 & $1$ & $1$ & $1.0$ & $\pm$ & $0.0$\\
12 & $2$ & $2$ & $2.0$ & $\pm$ & $0.0$\\
13 & $5$ & $5$ & $5.0$ & $\pm$ & $0.0$\\
14 & $15$ & $15$ & $15.0$ & $\pm$ & $0.0$\\
15 & $54$ & $54$ & $54.0$ & $\pm$ & $0.2$\\
16 & $167$ & $167$ & $169.0$ & $\pm$ & $1.7$\\
17 & $442$ & $447$ & $459.4$ & $\pm$ & $5.0$\\
18 & $987$ & $1022$ & $1096.6$ & $\pm$ & $13.7$\\
19 & $2032$ & $2174$ & $2625.7$ & $\pm$ & $38.4$\\
20 & $3363$ & $3737$ & $5717.8$ & $\pm$ & $102.4$\\
21 & $4675$ & $5397$ & $11545.1$ & $\pm$ & $247.6$\\
22 & $5776$ & $6848$ & $27115.7$ & $\pm$ & $2206.3$\\
23 & $6709$ & $8130$ & $86269.5$ & $\pm$ & $19246.7$\\
24 & $7674$ & $9506$ & $266233.4$ & $\pm$ & $6904.4$\\
25 & $8720$ & $10968$ & $840552.5$ & $\pm$ & $22760.4$\\
26 & $9525$ & $12141$ & $2448866.9$ & $\pm$ & $74150.6$\\
27 & $10106$ & $13027$ & $7540593.9$ & $\pm$ & $516159.0$\\
28 & $10386$ & $13463$ & $28705043.4$ & $\pm$ & $3557900.6$\\
29 & $10489$ & $13629$ & $169072346.1$ & $\pm$ & $52476936.5$\\
30 & $10535$ & $13698$ & $413482640.9$ & $\pm$ & $99900848.7$\\
\enddata
\tablecomments{Columns:
upper bound absolute magnitude $H$,
cumulative number of NEAs observed during the baseline survey period up to August 2014 \citep{2016AJ....151...80T},
cumulative number of known NEAs as of February 2016,
and full population as estimated in this manuscript (raw numbers).
The \emph{dip} in the nominal uncertainty at $H=24$ is due to the switching of binning from fine to coarse.
}
\label{tab:pop_data}
\end{deluxetable}

\clearpage

















\begin{thebibliography}{}
\bibitem[Anderson(2004)]{BOINC} Anderson, D.~P.\ 2004.\ BOINC: a system for public-resource computing and storage.\ Proceedings of the Fifth IEEE/ACM International Workshop on Grid Computing (GRID’04), 4-10.
\bibitem[Bottke et al.(1994)]{1994Icar..107..255B} Bottke, W.~F., Nolan, M.~C., Greenberg, R., Kolvoord, R.~A.\ 1994.\ Velocity distributions among colliding asteroids.\ Icarus 107, 255-268. 
\bibitem[Bottke et al.(2002)]{2002Icar..156..399B} Bottke, W.~F., Morbidelli, A., Jedicke, R., Petit, J.-M., Levison, H.~F., Michel, P., Metcalfe, T.~S.\ 2002.\ Debiased Orbital and Absolute Magnitude Distribution of the Near-Earth Objects.\ Icarus 156, 399-433. 
\bibitem[Brown et al.(2013)]{2013Natur.503..238B} Brown, P.~G., and 32 colleagues 2013.\ A 500-kiloton airburst over Chelyabinsk and an enhanced hazard from small impactors.\ Nature 503, 238-241. 
\bibitem[Granvik et al.(2016)]{2016Natur.530..303G} Granvik, M., Morbidelli, A., Jedicke, R., Bolin, B., Bottke, W.~F., Beshore, E., Vokrouhlick{\'y}, D., Delb{\`o}, M., Michel, P.\ 2016.\ Super-catastrophic disruption of asteroids at small perihelion distances.\ Nature 530, 303-306. 
\bibitem[Harris and D'Abramo(2015)]{2015Icar..257..302H} Harris, A.~W., D'Abramo, G.\ 2015.\ The population of near-Earth asteroids.\ Icarus 257, 302-312. 
\bibitem[Jedicke et al.(2016)]{2016Icar..266..173J} Jedicke, R., Bolin, B., Granvik, M., Beshore, E.\ 2016.\ A fast method for quantifying observational selection effects in asteroid surveys.\ Icarus 266, 173-188. 
\bibitem[Kessler(1981)]{1981Icar...48...39K} Kessler, D.~J.\ 1981.\ Derivation of the collision probability between orbiting objects The lifetimes of Jupiter's outer moons.\ Icarus 48, 39-48. 
\bibitem[Mainzer et al.(2011)]{2011ApJ...743..156M} Mainzer, A., and 36 colleagues 2011.\ NEOWISE Observations of Near-Earth Objects: Preliminary Results.\ The Astrophysical Journal 743, 156. 
\bibitem[Manley et al.(1998)]{1998A&AS..133..437M} Manley, S.~P., Migliorini, F., Bailey, M.~E.\ 1998.\ An algorithm for determining collision probabilities between small solar system bodies.\ Astronomy and Astrophysics Supplement Series 133, 437-444. 
\bibitem[Morbidelli et al.(2002)]{2002Icar..158..329M} Morbidelli, A., Jedicke, R., Bottke, W.~F., Michel, P., Tedesco, E.~F.\ 2002.\ From Magnitudes to Diameters: The Albedo Distribution of Near Earth Objects and the Earth Collision Hazard.\ Icarus 158, 329-342. 
\bibitem[Stuart and Binzel(2004)]{2004Icar..170..295S} Stuart, J.~S., Binzel, R.~P.\ 2004.\ Bias-corrected population, size distribution, and impact hazard for the near-Earth objects.\ Icarus 170, 295-311. 
\bibitem[Tricarico(2016)]{2016AJ....151...80T} Tricarico, P.\ 2016.\ Detection Efficiency of Asteroid Surveys.\ The Astronomical Journal 151, 80. 
\end{thebibliography}
\end{document}